# Transient Free-Surface Waves due to Impulsive Motion of a Submerged Source

JIAN-JUN SHU

School of Mechanical & Aerospace Engineering,

Nanyang Technological University, 50 Nanyang Avenue, Singapore 639798

e-mail: mjjshu@ntu.edu.sg

## Abstract

The problem of a viscous flow past a submerged source starting from rest and moving with a constant velocity, below and parallel to a free surface is examined. Asymptotic expressions for long-time evolution of free-surface elevation are obtained. The results show explicitly the viscous effect on the free-surface elevation and no surface tension effect on the asymptotic wave.

## 1 Introduction

The impulsive motion of a submerged source generates transient waves on the free surface of a viscous fluid. Lord Kelvin [1] ignored the free-surface shear stress and developed a theory to determine the kinematics and dynamics of the steady free-surface waves generated by a moving source with constant velocity in an inviscid fluid of infinite depth. It was found that the steady free-surface wave pattern consists of a series of so-called diverging waves and transverse waves. The diverging waves spread on each side of the moving source at an acute angle in relation to the source's moving direction, whereas the transverse waves move in the same direction as the moving source. The two wave systems intersect along the so-called "cusp locus" on both sides of the moving source. The angle between this line and the source's moving direction can be calculated as $19°28'$. The lines constitute the outer edge of the so-called Kelvin wake. It was also deduced that the diverging waves have a propagation direction of $35°16'$ compared to the moving direction at a certain distance from the source's navigation route. Havelock [2] studied the long-time evolution of a free surface due to an impulsive motion of a circular cylinder with a constant velocity in an inviscid fluid. Wehausen & Laitone [3], Chan & Chwang [4] and the author [5] enriched the theory with the consideration of viscous effect.

In the present paper the analytical approach will be extended to the transient interaction of impulsive motion of a submerged source with a free surface. It is interesting to study the viscous and surface tension effects on a point force moving beneath a free surface. The point force solution can be used to construct solutions for a submerged source of any shape.

## 2 Mathematical formulation

It is considered that a point force is submerged in a viscous incompressible fluid that occupies initially the lower half space $z < 0$ in a fixed Cartesian coordinate system. The point force is located at a distance $z_0$ below the free surface of the fluid, being suddenly started from rest and made to move with uniform velocity $U^* \vec{e}_x$, where $\vec{e}_x$ denotes the unit vector along the $x$ direction. It is meaningful to nondimensionalize the time by $\dfrac{U^*}{g}$, the distance by $\dfrac{U^{*2}}{g}$, the characteristic speed by $U^*$ and the pressure by $\rho U^{*2}$, where $g$ is the gravitational constant and $\rho$ the density of the fluid. Because only the wave profile at large distances downstream for high Reynolds numbers is investigated, the nondimensional, linearized Navier-Stokes equations describing the fluid flow, induced by an external body force with dimensionless strength $\vec{F} H(t)\, \delta(\vec{x} - \vec{x}_0)$, are given by

$$\nabla \cdot \vec{u}^* = 0, \qquad (1)$$

$$\frac{\partial \vec{u}^*}{\partial t} + \frac{\partial \vec{u}^*}{\partial x} = -\nabla p^* + \varepsilon \nabla^2 \vec{u}^* + \vec{F} H(t)\, \delta(\vec{x} - \vec{x}_0). \qquad (2)$$

At negative time $t < 0$ everything is at rest,

$$\vec{u}^* = \vec{0}, \quad p^* = 0, \quad \eta = 0, \quad \text{for} \quad t < 0. \qquad (3)$$

On the free surface, the tangential stresses are assumed to vanish and the normal stress has a jump proportional to the surface tension and mean curvature. On $z = 0$ these free-surface conditions are linearized together with the kinematic boundary condition to yield

$$\frac{\partial u^*}{\partial z} + \frac{\partial w^*}{\partial x} = 0, \quad \frac{\partial v^*}{\partial z} + \frac{\partial w^*}{\partial y} = 0, \qquad (4)$$

$$\eta - p^* + 2\varepsilon \frac{\partial w^*}{\partial z} = \sigma\left(\frac{\partial^2}{\partial x^2} + \frac{\partial^2}{\partial y^2}\right)\eta, \tag{5}$$

$$\left(\frac{\partial}{\partial t} + \frac{\partial}{\partial x}\right)\eta = w^*. \tag{6}$$

The fluid velocity and pressure vanish at infinity,

$$\vec{u}^* \to \vec{O}, \quad p^* \to 0, \quad \text{as} \quad z \to -\infty. \tag{7}$$

Here the variables $\vec{u}^* = (u^*, v^*, w^*)^T$ and $p^*$ represent the non-dimensional disturbed velocity and disturbed pressure in the fluid, $\vec{x} = (x, y, z)^T$ and $\vec{x}_0 = (0, 0, -z_0)^T$ are the field point and source point, $\eta$ is the free-surface elevation, and the dimensionless parameters $\varepsilon = \frac{1}{R_e}$ and $\sigma = \frac{1}{W_e}$ can be regarded as the reciprocal of the Reynolds number $R_e = \frac{\rho U^{*3}}{g \times (\text{dynamic viscosity})}$ and the Weber number $W_e = \frac{\rho U^{*4}}{g \times (\text{surface tension})}$ respectively. $H(\bullet)$ and $\delta(\bullet)$ are Heaviside's step function and Dirac delta function respectively. The solution to Eqs. (1) and (2) for an unbounded fluid domain is given by the author [6] as

$$\vec{u}_0 = -\frac{H(t)}{4\pi}\vec{F} \bullet \left(\mathbf{I}\nabla^2 - \nabla\nabla\right)\int_0^t \frac{\mathrm{erf}\left(r^*/2\sqrt{\varepsilon\tau}\right)}{r^*}d\tau, \quad p_0 = \frac{H(t)\vec{F}\bullet\vec{x}}{4\pi r^3}, \tag{8}$$

where $\vec{x}^* = \vec{x} - \tau\vec{e}_x$, $r^* = \|\vec{x}^* - \vec{x}_0\|$, and $r = \|\vec{x} - \vec{x}_0\|$.

Now let the entire solution be written as

$$\vec{u}^* = \vec{u}_0 + \vec{u}, \quad p^* = p_0 + p. \tag{9}$$

To reduce the number of variables involved, the motion can be represented as a potential flow plus a rotational flow. Thus, two new functions $\phi$ and $\vec{\omega} = (\omega_x, 0, \omega_z)^T$ are defined§ by

$$\vec{u} = \nabla\phi + \nabla \times \vec{\omega} \tag{10}$$

such that

$$\nabla^2 \phi = 0, \tag{11}$$

$$\frac{\partial \vec{\omega}}{\partial t} + \frac{\partial \vec{\omega}}{\partial x} = \varepsilon \nabla^2 \vec{\omega}, \tag{12}$$

---

§ As three new independent variables, $\phi$, $\omega_x$ and $\omega_z$ replace $u$, $v$ and $w$ (three components of $\vec{u}$), the three independent replacements only allow $\vec{\omega}$ having two independent components. Without loss of generality, say $\omega_y = 0$.

and

$$p = -\frac{\partial \phi}{\partial t} - \frac{\partial \phi}{\partial x} \tag{13}$$

subject to the condition $\eta = 0$, for $t < 0$. By means of the Laplace transform in $t$ and the Fourier transforms in $x$ and $y$,

$$[\bar{\phi}, \bar{\vec{\omega}}](s, \alpha, \beta, z) = \int_0^\infty \int_{-\infty}^\infty \int_{-\infty}^\infty [\phi, \vec{\omega}](t, x, y, z) \exp(-st - i\alpha x - i\beta y - [A, B](z + z_0)) \, dt \, dx \, dy, \tag{14}$$

$$\bar{\eta}(s, \alpha, \beta) = \int_0^\infty \int_{-\infty}^\infty \int_{-\infty}^\infty \eta(t, x, y) \exp(-st - i\alpha x - i\beta y) \, dt \, dx \, dy, \tag{15}$$

it then follows from Eqs. (11) to (13) that $A$ and $B$ must satisfy

$$A = \sqrt{\alpha^2 + \beta^2}, \quad B = \sqrt{\alpha^2 + \beta^2 + \frac{s + i\alpha}{\varepsilon}}. \tag{16}$$

The free-surface conditions (4) to (6) can be expressed in terms of $\bar{\phi}$, $\bar{\omega}_x$, $\bar{\omega}_z$ and $\bar{\eta}$ as

$$\mathbf{C}\vec{V} = \vec{C}^{\{A\}} \exp(-A(z + z_0)) + \vec{C}^{\{B\}} \exp(-B(z + z_0)) \tag{17}$$

where superscripts $\{A\}$ and $\{B\}$ denote contributions by the potential flow and the rotational flow respectively, $\vec{V} = (\bar{\phi}, \bar{\omega}_x, \bar{\omega}_z, \bar{\eta})^T$,

$$\mathbf{C} = \begin{bmatrix} 2i\alpha A(s+i\alpha)\varepsilon^{3/2} & \alpha\beta(s+i\alpha)\varepsilon^{3/2} & i\beta B(s+i\alpha)\varepsilon^{3/2} & 0 \\ 2i\beta A(s+i\alpha)\varepsilon^{3/2} & (\beta^2+B^2)(s+i\alpha)\varepsilon^{3/2} & -i\alpha B(s+i\alpha)\varepsilon^{3/2} & 0 \\ (s+i\alpha)^2 + A[1+A^2\sigma+2A(s+i\alpha)\varepsilon] & -i\beta[1+A^2\sigma+2B(s+i\alpha)\varepsilon] & 0 & 0 \\ A & -i\beta & 0 & -(s+i\alpha) \end{bmatrix} \tag{18}$$

and

$$\vec{C}^{\{A\}} = \frac{\alpha F_x + iAF_z}{2s(s+i\alpha)A} \begin{bmatrix} -2\alpha A(s+i\alpha)\varepsilon^{3/2} \\ -2\beta A(s+i\alpha)\varepsilon^{3/2} \\ i[A - (s+i\alpha)^2 + A^3\sigma - 2A^2(s+i\alpha)\varepsilon] \\ 0 \end{bmatrix}, \tag{19}$$

$$\vec{C}^{\{B\}} = -\frac{i\alpha(1 + A^2\sigma)F_x}{2s(s+i\alpha)} \begin{bmatrix} 0 \\ 0 \\ 1 \\ 0 \end{bmatrix} + O(\sqrt{\varepsilon}). \tag{20}$$

The solution may be expressed as

$$\bar{\eta} = -\frac{i\left(\alpha F_x + iAF_z\right)\left[(s+i\alpha)^2 - A\left(1+A^2\sigma\right)\right]\exp(-Az_0)}{2sA\left[(s+i\alpha)^2 + A\left(1+A^2\sigma\right) + 4A^2(s+i\alpha)\varepsilon\right]} + O(\varepsilon^{3/2}). \tag{21}$$

Because of the exponential term $\exp(-B(z+z_0))$ involved in Eq. (17), the expansion of $\vec{C}^{\{B\}}$ in Eq. (20) will not affect the final result in Eq. (21).

## 3 Results

The wave number vector $(\alpha, \beta)^T$ identifies a free-surface wave in the far wake of a point force so long as it points to a curve in the $(\alpha, \beta)$-plane defined by setting one factor of the denominator in Eq. (21) to zero,

$$(s+i\alpha)^2 + A\left(1+A^2\sigma\right) + 4A^2(s+i\alpha)\varepsilon = 0. \tag{22}$$

The curve consists of three or two branches depending on whether $\|s\| < 1/4$ or $\|s\| \geq 1/4$. In the steady limit $(s=0)$, one closed branch shrinks to zero, while the other two open branches represent the Kelvin wave pattern [1], which consists of a series of so-called diverging waves and transverse waves.

The problem posed here is that of finding the far-field asymptotic behaviour of free-surface waves, induced by impulsive motion of a submerged point force $\vec{F} = (F_x, 0, F_z)$. $F_x$ and $F_z$ represent dimensionless drag and lift forces respectively. For the unsteady case, i.e., $\|s\| \neq 0$, the cylindrical coordinates $(R, \theta)$ are introduced on the free surface through

$$x = R\cos\theta, \quad y = R\sin\theta. \tag{23}$$

To obtain the leading terms in the far-field asymptotic representation for small $\varepsilon$ and $s$, Lighthill's two-stage scheme [7] shall be employed, which in essence involves calculating the $\alpha$-integration by residues [8] and the $R$-integration by the method of steepest descent [9].

For the first stage of Lighthill's scheme, the roots of the pole equation are considered

$$(s+i\alpha)^2 + A\left(1+A^2\sigma\right) + 4A^2(s+i\alpha)\varepsilon = 0. \tag{24}$$

For small $\varepsilon$ and $s$, the roots $\left(\alpha^{(j)}, j=1,2\right)$ take the form of

$$\alpha^{(j)} = (-1)^{j-1}A_2 - \frac{2iA_1^2\left(s+2A_1^2\varepsilon\right)}{3A_2^2 - 2A_1(A_1+1)} + O\left(\varepsilon^{3/2} + \varepsilon s + s^2\right), \tag{25}$$

where $A_2 = \sqrt{A_1\left(1+A_1^2\sigma\right)}$ and $A_1$ satisfies a cubic equation $\sigma A_1^3 - A_1^2 + A_1 + \beta^2 = 0$, that is,

$$A_1 = \begin{cases} \dfrac{1}{3\sigma}\left[1+2\sqrt{1-3\sigma}\cos\left(\dfrac{\pi+\psi}{3}\right)\right], \quad \cos\psi = \dfrac{27\beta^2\sigma^2+9\sigma-2}{2(1-3\sigma)^{3/2}}, \quad 0\le\psi\le\pi \quad \text{if} \quad \sigma\le\dfrac{1}{4} \\ \text{no real positive root} \hspace{24em} \text{if} \quad \sigma>\dfrac{1}{4} \end{cases} \quad (26)$$

Using the residue theorem of a meromorphic function with respect to $\alpha$, the leading terms contributing significantly to the asymptotic expressions about $\varepsilon=0$ and $s=0$ of the free-surface elevation can be written as

$$\int_0^\infty \eta\exp(-st)dt = -\frac{1}{2\pi s}\sum_{j=1}^2 \int_{-\infty}^\infty \frac{\left[A_2 F_x + (-1)^{j-1} i A_1 F_z\right] A_1 A_2}{3 A_2^2 - 2 A_1 (A_1+1)} \exp(-A_1 z_0 + i R h_j)[1+O(\varepsilon+s)]\,d\beta, \tag{27}$$

where $h_j(\beta/\alpha^{(j)}) = \alpha^{(j)}\cos\theta + \beta\sin\theta$.

For the second stage of Lighthill's scheme, the saddle points are considered to satisfy the derivative of the exponent of the Fourier kernel,

$$\frac{\partial}{\partial\beta} h_j(\beta/\alpha^{(j)}) = \frac{\partial\alpha^{(j)}}{\partial\beta}\cos\theta + \sin\theta = 0. \tag{28}$$

For small $\varepsilon$ and $s$, the roots $\beta_\pm^{(j)}$ take the form of

$$\beta_\pm^{(j)} = (-1)^{j-1}\frac{3A_4^2 - 2(A_3+1)A_3}{3A_4^2 - 2A_3}\left\{1 + (-1)^{j-1}\frac{4i A_4^3\left[(3A_4^2 - 4A_3)s - 2(3A_4^2 - 4A_3)A_3^2\varepsilon\right]}{3A_4^6 - 6A_3^2 A_4^4 + 14 A_3 A_4^4 - 28 A_3^2 A_4^2 + 8 A_3^4 + 8 A_3^3}\right\}$$
$$\times A_4 \tan\theta + O(\varepsilon^{3/2} + \varepsilon s + s^2), \tag{29}$$

where $A_4 = \sqrt{A_3(1 + A_3^2 \sigma)}$ and $A_3$ satisfies a sextic equation

$$(1 + A_3^2\sigma)(3A_3^2\sigma - 2A_3 + 1)^2 \tan^2\theta + (A_3^2\sigma - A_3 + 1)(3A_3^2\sigma + 1)^2 = 0. \tag{30}$$

The sextic equation (30) cannot be solved by rational operations and root extraction on coefficients. For small $\sigma$, the root $A_3$ takes the form of

$$A_3 = M_\pm\left[1 + \frac{(5G_\pm - 6)M_\pm^2\sigma}{G_\pm - 2} + O(\sigma^2)\right] \tag{31}$$

with the $G_\pm$ and $M_\pm$ written as

$$M_\pm = \frac{G_\pm + 1}{2}, \quad G_\pm = \frac{1 \pm \sqrt{1 - 8\tan^2\theta}}{4\tan^2\theta}. \tag{32}$$

Hence

$$A_1 = M_{\pm} + (-1)^j \frac{4i(G_{\pm}-1)M_{\pm}^{1/2} s}{G_{\pm}(G_{\pm}-2)} + (-1)^{j-1} \frac{8i(G_{\pm}-1)(2G_{\pm}-1)M_{\pm}^{5/2} \varepsilon}{G_{\pm}(G_{\pm}-2)}$$
$$+ \frac{(5G_{\pm}-6)M_{\pm}^{3} \sigma}{G_{\pm}-2} + O(\varepsilon^{3/2} + \varepsilon\sigma + \sigma^2 + \varepsilon s + \sigma s + s^2), \tag{33}$$

$$A_2 = M_{\pm}^{1/2} + (-1)^j \frac{2i(G_{\pm}-1)s}{G_{\pm}(G_{\pm}-2)} + (-1)^{j-1} \frac{4i(G_{\pm}-1)(2G_{\pm}-1)M_{\pm}^{2} \varepsilon}{G_{\pm}(G_{\pm}-2)}$$
$$+ \frac{(3G_{\pm}-4)M_{\pm}^{5/2} \sigma}{G_{\pm}-2} + O(\varepsilon^{3/2} + \varepsilon\sigma + \sigma^2 + \varepsilon s + \sigma s + s^2), \tag{34}$$

$$A_4 = M_{\pm}^{1/2}\left[1 + \frac{(3G_{\pm}-4)M_{\pm}^{2} \sigma}{G_{\pm}-2} + O(\sigma^2)\right], \tag{35}$$

$$\alpha_{\pm}^{(j)} = (-1)^{j-1} M_{\pm}^{1/2}\left\{1 + (-1)^{j-1} \frac{i(G_{\pm}-3)s}{(G_{\pm}-2)M_{\pm}^{1/2}} + (-1)^{j-1} \frac{2i(5G_{\pm}-7)M_{\pm}^{3/2} \varepsilon}{G_{\pm}-2}\right.$$
$$\left.+ \frac{(3G_{\pm}-4)M_{\pm}^{2} \sigma}{G_{\pm}-2}\right\} + O(\varepsilon^{3/2} + \varepsilon\sigma + \sigma^2 + \varepsilon s + \sigma s + s^2), \tag{36}$$

$$\beta_{\pm}^{(j)} = (-1)^{j} G_{\pm} M_{\pm}^{1/2}\left\{1 + (-1)^{j} \frac{4is}{(G_{\pm}-2)M_{\pm}^{1/2}} + (-1)^{j-1} \frac{8i(2G_{\pm}-1)M_{\pm}^{3/2} \varepsilon}{G_{\pm}-2}\right.$$
$$\left.+ \frac{(5G_{\pm}-2)M_{\pm}^{2} \sigma}{G_{\pm}-2}\right\}\tan\theta + O(\varepsilon^{3/2} + \varepsilon\sigma + \sigma^2 + \varepsilon s + \sigma s + s^2). \tag{37}$$

Using the method of steepest descent, the leading term is obtained as

$$\int_0^{\infty} \eta \exp(-st)dt = \frac{1}{s}\sqrt{\frac{1}{2\pi R\cos\theta}}(1-8\tan^2\theta)^{-1/4}\sum_{j=1}^{2}\sum_{\pm} M_{\pm}^{-1/4}\left[M_{\pm} F_x + (-1)^{j-1} iM_{\pm}^{3/2} F_z\right]$$
$$\times\left\{\exp\left(-M_{\pm} z_0 + (-1)^{j-1} iR\frac{M_{\pm}^{3/2}}{G_{\pm}}\left[1+(-1)^{j-1} 4iM_{\pm}^{3/2}\varepsilon + M_{\pm}^{2}\sigma\right]\cos\theta \pm (-1)^j i\frac{\pi}{4}\right)\right. \tag{38}$$
$$\left.+ O\left(\varepsilon + \sigma + s + \frac{1}{R}\right)\right\} \times \exp\left(-\frac{2RM_{\pm}\cos\theta}{G_{\pm}}s\right).$$

Upon substituting and other mathematical manipulations, the free-surface elevation can formally be expressed as

$$\int_0^{\infty} \eta \exp(-st)dt = \frac{1}{s}\sqrt{\frac{2}{\pi R\cos\theta}}(1-8\tan^2\theta)^{-1/4}\sum_{\pm}\left[M_{\pm}^{3/4} P_{\pm}\left(F_x \cos\gamma_{\pm} - M_{\pm}^{1/2} F_z \sin\gamma_{\pm}\right)\right.$$
$$\left.+ O\left(\varepsilon + \sigma + s + \frac{1}{R}\right)\right] \times \exp\left(-\frac{2RM_{\pm}\cos\theta}{G_{\pm}}s\right) \tag{39}$$

where

$$P_\pm = \exp\left(-\frac{M_\pm\left(z_0\, G_\pm + 4R M_\pm^2\, \varepsilon \cos\theta\right)}{G_\pm}\right), \quad \gamma_\pm = \frac{R M_\pm^{3/2} \cos\theta}{G_\pm} \mp \frac{\pi}{4}. \tag{40}$$

Using the inverse Laplace transform with respect to $s$, the asymptotic expression of the free-surface elevation can be written as

$$\eta = \sqrt{\frac{2}{\pi R \cos\theta}} \left(1 - 8\tan^2\theta\right)^{-1/4} \sum_\pm M_\pm^{3/4} P_\pm \left(F_x \cos\gamma_\pm - M_\pm^{1/2} F_z \sin\gamma_\pm\right)$$
$$\times H\left(t - \frac{2R M_\pm \cos\theta}{G_\pm}\right) + O\left(\varepsilon + \sigma + \frac{1}{R}\right). \tag{41}$$

## 4  Conclusions

The Taylor expansion has been performed for the transient free-surface flow due to impulsive motion of a submerged point force. The new asymptotic expressions of free-surface elevations and wavefronts for large time are obtained including the effects of viscosity and surface tension. The presence of viscosity is found to reduce the free-surface wave amplitude, while the surface tension does not affect the asymptotic wave.

It is worth to mention to this end that the present new result (41) is dealing with the Oseen problem [6] of large Reynolds number $\left(\varepsilon = \frac{1}{R_e}\right)$ and far field (asymptotic wave), in contrast to the classical result [10] based on unsteady Stokes solution, which is only valid for small Reynolds numbers and near field around the submerged source.